\documentclass[floatfix,aps,prl,showpacs,amsmath,nofootinbib,preprintnumbers,twocolumn]{revtex4}

\usepackage{graphicx}
\usepackage{bm}

\newcommand{\figurewidth}{2.3in}
\newcommand{\figurewidthL}{2.3in}

\def\({\left(}
\def\){\right)}
\def\[{\left[}
\def\]{\right]}

\def\e{\begin{equation}}
\def\q{\end{equation}}
\def\m{\begin{eqnarray}}
\def\n{\end{eqnarray}}

\begin{document}

\title{The Tilt of Primordial Gravitational Waves Spectra from BICEP2}

\author{Cheng Cheng and Qing-Guo Huang}\email{huangqg@itp.ac.cn}
\affiliation{State Key Laboratory of Theoretical Physics, Institute of Theoretical Physics, Chinese Academy of Science, Beijing 100190, People's Republic of China}

\date{\today}

\begin{abstract}

In this paper we constrain the tilt of the spectra of primordial gravitational waves from Background Imaging of Cosmic Extragalactic Polarization (BICEP2) data only. We find $r=0.21_{-0.10}^{+0.04}$ and $n_t=-0.06_{-0.23}^{+0.25}$ (at $68\%$ C.L.) which implies that a scale-invariant primordial gravitational waves spectra is consistent with BICEP2 nicely. Our results  provide strong evidence for supporting inflation model, and the alternative models, for example the ekpyrotic model which predicts $n_t=2$, are ruled out at more than $5\sigma$ significance.

\end{abstract}


\maketitle


Hint for the primordial gravitational waves was illustrated from the combination of Planck TT \cite{Ade:2013zuv} and Wilkinson Microwaves Anisotropy Probe (WMAP) 9-year TE data \cite{Hinshaw:2012aka} at very large scales (i.e. $\ell\leq 100$) in \cite{Zhao:2014rna} where the tensor-to-scalar ratio $r>0$ is preferred at more than $68\%$ C.L. and the maximized likelihood value of $r$ is around 0.2 which confirms the previous one in \cite{Zhao:2010ic}. 
Recently the breakthrough in hunting for the signal of relic gravitational waves was reported by Background Imaging of Cosmic Extragalactic Polarization (BICEP2) \cite{Ade:2014xna}, namely 
\m
r=0.20_{-0.05}^{+0.07}, 
\label{rb14}
\n
with $r=0$ disfavored at $7.0\sigma$.
It must be one of the most important discoveries in this century and a new era of cosmology is coming.  

The next important task is to explore the property of relic gravitational waves, for example the tilt of its spectra which is defined by 
\m
n_t\equiv {d\ln P_t\over d\ln k}, 
\n
or equivalently,  
\m
P_t(k)=P_t(k_p)\({k\over k_p}\)^{n_t}, 
\label{pt}
\n
where $P_t$ is the amplitude of the relic gravitational waves spectra and $k_p=0.004$ Mpc$^{-1}$ is the pivot scale. BICEP2 finds an excess of B-mode power over the base lensed-$\Lambda$CDM expectation in the range of $\ell\in [30,150]$ multipoles. Since this range does not cover broad perturbation modes, the power-law spectra of primordial gravitational waves in Eq.~(\ref{pt}) is expected to be good enough. In this paper $n_t$ is taken as a free parameter which can be constrained by using the recent data released by BICEP2 \cite{Ade:2014xna}.

On the other hand, inflation \cite{Guth:1980zm,Linde:1981mu,Albrecht:1982wi} is an elegant paradigm for the early Universe. Not only does the inflation solve the flatness and horizon problems in the hot big bang model, but also the quantum fluctuations generated during inflation provide tiny primordial inhomogeneities to seed the anisotropies in the Cosmic Microwave Background (CMB) radiation and formation of large-scale structure. The predictions of inflation, such as the flatness and near scale-invariant adiabatic scalar perturbations, have been confirmed by WMAP \cite{Hinshaw:2012aka}, Planck \cite{Ade:2013zuv} and some other highL CMB data including Atacama Cosmology Telescope (ACT) \cite{Sievers:2013ica} and South Pole Telescope (SPT) \cite{Story:2012wx}. Now another important prediction of inflation, relic gravitational waves, has also been detected by BICEP2 \cite{Ade:2014xna}. 

However there are also some alternative models for the early Universe, for example the ekpyrotic model \cite{Khoury:2001wf} which predicts a very blue tilted spectra of primordial gravitational waves, namely $n_t=2$. 
But the simplest version of inflation, so-called canonical single-field slow-roll inflation, predicts a consistency relation between $n_t$ and $r$: $r=-8 n_t$ \cite{Liddle:1992wi}. For $r\simeq 0.2$, $n_t\simeq -0.025$, or equivalently the spectra of relic gravitational waves is nearly scale-invariant as well.  For the more general single-field inflation, the consistency relation becomes $r=-8 c_s n_t$ \cite{Garriga:1999vw}, where $c_s$ is the sound speed for the scalar perturbation modes, and the canonical single-field slow-roll inflation is recovered when $c_s=1$. If $c_s$ is not too small, a nearly scale-invariant spectra of relic gravitational waves is still expected. Actually in the inflation model, $n_t=-2\epsilon$ \cite{Liddle:1992wi,Garriga:1999vw} and inflation requires $\ddot a/a=H^2(1-\epsilon)>0$, and hence $-2<n_t\leq 0$, where $\epsilon\equiv \dot H/H^2$.

\begin{figure*}[hts]
\includegraphics[width=\figurewidth]{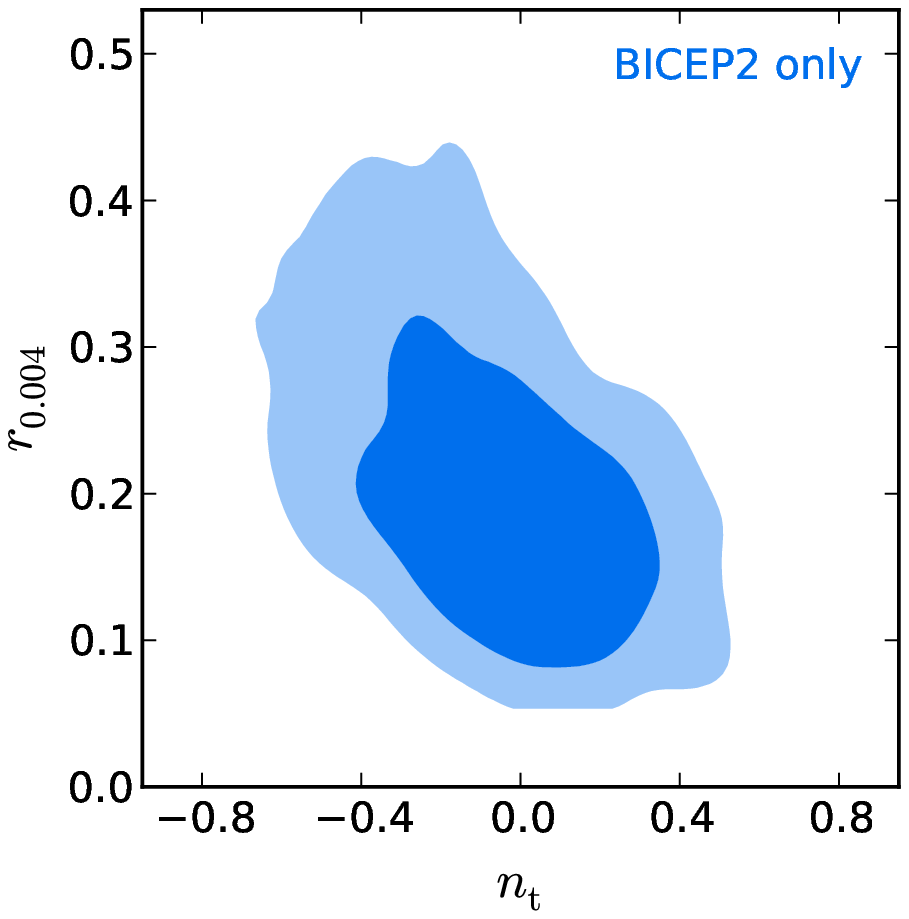}
\includegraphics[width=\figurewidthL]{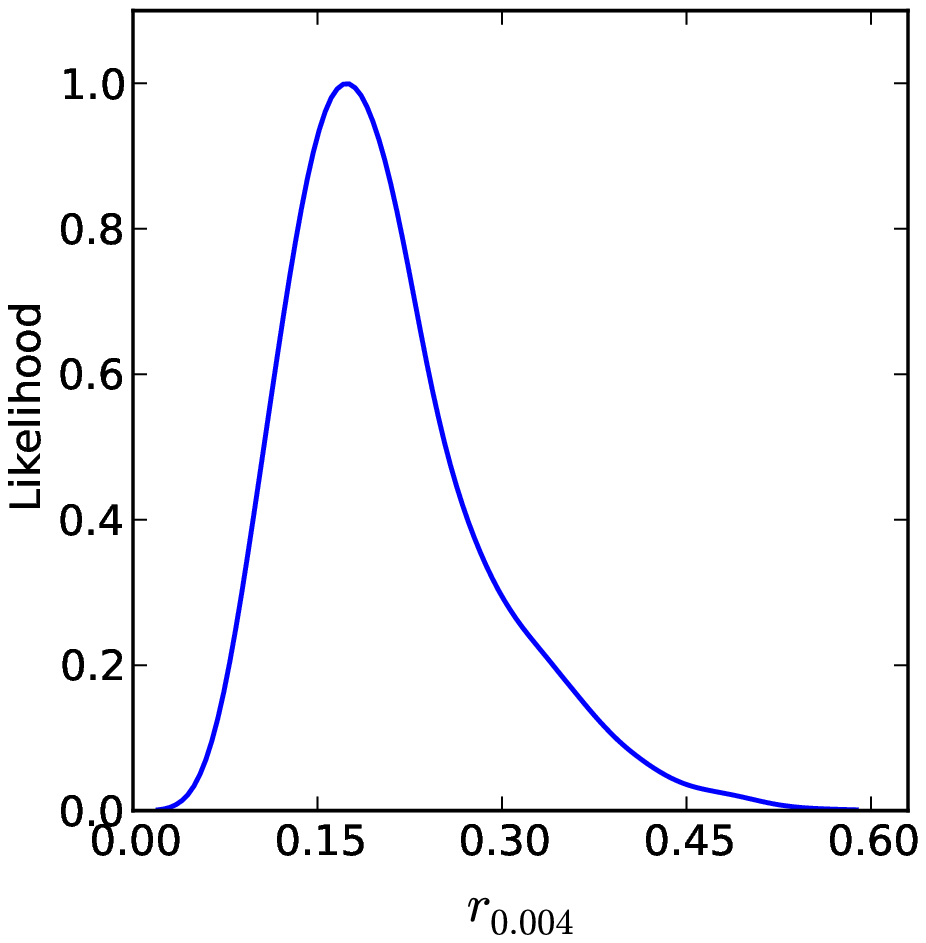}
\includegraphics[width=\figurewidthL]{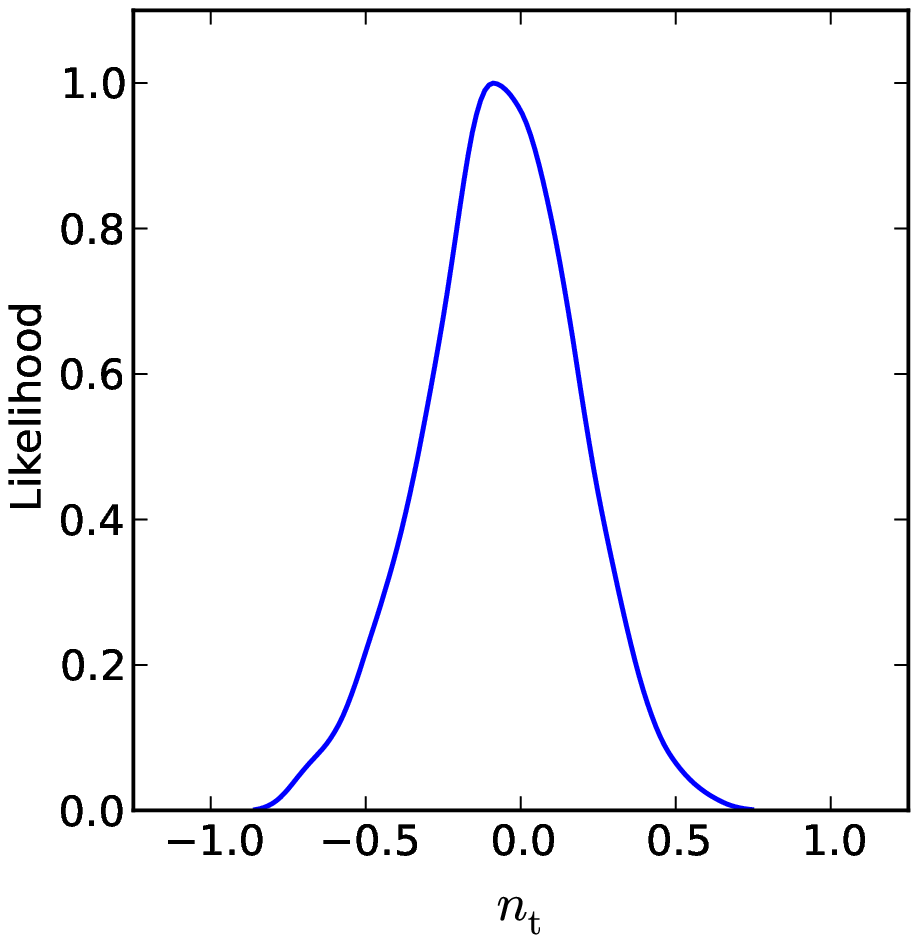}
\caption{The constraints on $r$ and $n_t$ from BICEP2 data only. The left panel is the contour plot of $r$ and $n_t$, the middle and right panels show the likelihood distributions of $r$ and $n_t$ respectively. 
}
\label{fig:brnt}
\end{figure*}

In this paper we do not assume any model and the tilt $n_t$ is taken as a fully free parameter. We consider uniform priors for the tensor-to-scalar ratio $r$ and tilt $n_t$ in the ranges of $r\in [0, 2]$ and $n_t\in [-4, 4]$ respectively.  Using the BICEP2 data \cite{Ade:2014xna} only, we run CosmoMC \cite{cosmomc} to figure out the constraints on $r$ and $n_t$ as follows 
\m
r=0.21_{-0.10}^{+0.04},\quad n_t=-0.06_{-0.23}^{+0.25}
\n
at $68\%$ C.L. 
respectively. A scale-invariant spectra of primordial gravitational waves is consistent with BICEP2 data. The contour plot of $r$ and $n_t$ and the likelihood distributions for each of them shows up in Fig.~\ref{fig:brnt}.
From Fig.~\ref{fig:brnt}, we see that $n_t=2$ is ruled out at more than $5\sigma$ level. Our results indicate that the ekpyrotic model has been significantly ruled out with high confidence level. The canonical single-field slow-roll inflation is within $1\sigma$ error. But the error bar of $n_t$ is still quite big, and we hope that the consistency relation of canonical single-field slow-roll inflation can be carefully tested in the future.

In a word, our results provide strong evidence for inflation, and the alternative models for the early Universe, for example the ekpyrotic model, are ruled out at high confidence level. 
Detection of the relic gravitational waves opens a new window to explore the physics in the early Universe and a lot of investigations shall be done in the near future.

Before closing this letter, we also want to mention that combining with WMAP Polarization and other highL CMB data, Planck implies a much smaller tensor-to-scalar ratio $r<0.11$ (at $95\%$ C.L.) \cite{Ade:2013zuv} if a power-law spectra of adiabatic scalar perturbations is assumed. A similar results with $r<0.12$ (at $95\%$ C.L.) \cite{Cheng:2013iya} was obtained from the combination of WMAP 9-year data \cite{Hinshaw:2012aka}, highL CMB data, Baryon Acoustic Oscillation (BAO) \cite{BAO} and $H_0$ prior from Hubble Space Telescope (HST) \cite{Riess:2011yx}. We see that there is a moderately strong tension on $r$ between BICEP2 and other CMB data in the $\Lambda$CDM+tensors model. A simple way to relax this tension is to take the running of spectral index into account \cite{Ade:2014xna}. A careful analysis will be done elsewhere \cite{Cheng:2014s}. Surprisingly, before Planck released its data, we also observed a peak around $r\sim 0.1$ in the one-dimensional marginalized likelihood distribution (see the dotted curve in Figure.~2 of \cite{Cheng:2013iya} when the running of spectral index $\alpha_s\equiv dn_s/d\ln k$ was considered) and a blue tilted power spectra of adiabatic scalar perturbations is preferred at large scales which is consistent with our results in \cite{Zhao:2014rna}. We believe that it is worthy hunting for some hints or signals of relic gravitational waves from WMAP and Planck data in the near future, and then we can approach to a better understanding about the physics in the early Universe. 


\vspace{5mm}
\noindent {\bf Acknowledgments. }
We acknowledge the use of Planck Legacy Archive, ITP and Lenovo
Shenteng 7000 supercomputer in the Supercomputing Center of CAS
for providing computing resources. This work is supported by the project of Knowledge Innovation Program of Chinese Academy of Science and grants from NSFC (grant NO. 10821504, 11322545 and 11335012).



\end{document}